\documentclass[12pt,draftclsnofoot, perreview, onecolumn]{IEEEtran}

\normalsize

\usepackage{amsfonts}
\usepackage{amsmath}
\usepackage{mathrsfs}
\usepackage{amssymb}
\usepackage{enumerate}
\usepackage{color,xcolor}
\usepackage{graphicx}
\usepackage{subfigure}
\usepackage{multirow,tabularx}
\usepackage{balance}
\usepackage{citesort}

\ifCLASSOPTIONonecolumn
\newcommand{\figwidth}{0.7\textwidth}
\fi
\ifCLASSOPTIONtwocolumn
\newcommand{\figwidth}{0.45\textwidth}
\fi

\newtheorem{algorithm}{\textbf{Algorithm}}
\newtheorem{example}{\textbf{Example}}

\makeatletter
    
    \newcommand{\Rmnum}[1]{\expandafter\@slowromancap\romannumeral #1@}
\makeatother

\begin{document}
\title{Unequal Error Protection by Partial Superposition Transmission Using LDPC Codes}
\author{Kechao~Huang, Chulong~Liang, Xiao~Ma,~\IEEEmembership{Member,~IEEE,} and~Baoming~Bai,~\IEEEmembership{Member,~IEEE}
\thanks{This work is supported by the 973 Program~(No.2012CB316100) and the NSF~(No.61172082) of China.}
\thanks{K.~Huang, C.~Liang, and~X.~Ma are with the Department of Electronics and Communication Engineering, Sun Yat-sen University, Guangzhou 510006, China~(e-mail:~hkech@mail2.sysu.edu.cn, lchul@mail2.sysu.edu.cn, maxiao@mail.sysu.edu.cn).}
\thanks{B.~Bai is with the State Key Lab.~of ISN, Xidian University, Xi'an 710071, China~(e-mail:~bmbai@mail.xidian.edu.cn).}
}

\maketitle

\begin{abstract}
In this paper, we consider designing low-density parity-check~(LDPC) coded modulation systems to achieve unequal error protection (UEP). We propose a new UEP approach by partial superposition transmission called UEP-by-PST. In the UEP-by-PST system, the information sequence is distinguished as two parts, the more important data~(MID) and the less important data~(LID), both of which are coded with LDPC codes. The codeword that corresponds to the MID is superimposed on the codeword that corresponds to the LID. The system performance can be analyzed by using discretized density evolution. Also proposed in this paper is a criterion from a practical point of view to compare the efficiencies of different UEP approaches. Numerical results show that, over both additive white Gaussian noise~(AWGN) channels and uncorrelated Rayleigh fading channels, 1)~UEP-by-PST provides higher coding gain for the MID compared with the traditional equal error protection~(EEP) approach, but with negligible performance loss for the LID; 2)~UEP-by-PST is more efficient with the proposed practical criterion than the UEP approach in the digital video broadcasting~(DVB) system.


\end{abstract}

\IEEEpeerreviewmaketitle

\section{Introduction}

\IEEEPARstart{I}{n} many practical communication systems such as wireless networks, control applications and interactive systems, data can be partitioned into several parts that have different degrees of significance. For example, in wireless communication system, headers of the medium access control (MAC) frame such as frame control, duration and address are more important than the frame body, because an error in the header may lead to the rejection of the frame while errors in the frame body are usually tolerable. Traditional equal error protection~(EEP) approach is usually not the most efficient way to guarantee the quality of the important data. Hence unequal error protection~(UEP) is required to make the best use of the resources~(say bandwidth).

A practical approach to achieving UEP is based on modulation. In~\cite{Wei93}, the author introduced a UEP approach based on a nonuniform arrangement of the signal constellation, also known as multiresolution modulation~\cite{Ramchandran93} or hierarchical modulation~\cite{Morimoto96}. In such a constellation, more important bits in a constellation symbol have larger minimum Euclidian distance than less important bits. In~\cite{Morelos00}, a UEP approach using uniformly spaced constellation was proposed, where different bits in a constellation symbol have different average number of nearest neighbors. However, these UEP approaches can achieve only a limited number of UEP levels for a given constellation. More recently, the authors of~\cite{Chang12} proposed a method of achieving arbitrarily large number of UEP levels by using multiplexed hierarchical quadrature amplitude modulation~(QAM) constellations.

An alternative approach to achieving UEP is based on channel coding. In this approach, more powerful error-correction coding is applied to the more important data (MID) than the less important data (LID). UEP codes were firstly introduced by Masnick~{\em et al} in 1967~\cite{Masnick67}. In~\cite{Lin90}, the authors found all the cyclic UEP codes of odd length up to 65 by computer searching. In~\cite{Hagenauer88}, a UEP approach using rate-compatible punctured convolutional (RCPC) codes was proposed whereby the more important bits were punctured less frequently than the less important bits. In~\cite{Barbulescu95}, turbo codes were employed for UEP in the same way as RCPC codes. Research on UEP low-density parity-check~(LDPC) codes can be found in~\cite{Rahnavard04,Rahnavard07,Kumar06}. In~\cite{Rahnavard04,Rahnavard07}, UEP LDPC codes were constructed by designing the variable node degree distribution of the code in an irregular way. In~\cite{Kumar06}, the authors proposed a new class of UEP LDPC codes based on Plotkin-type constructions. In order to provide more efficient UEP, error-correction coding and modulation can be jointly used~\cite{Aydinlik04,Aydinlik08}. These methods based on channel coding and/or modulation have been widely used for image and layered video transmission~\cite{DVB04,Barmada05,Deetzen08,Alajel12,Chung10,Zhang11}.

To the best of our knowledge, all the existing UEP approaches improve the performance of the MID by sacrificing the performance of the LID. Another issue is that no simple criteria were mentioned in the literatures to compare the efficiencies of different UEP approaches. In this paper, motivated by recent work on constructing long codes from short codes by block Markov superposition transmission~\cite{Ma_ISIT_13}, we propose a new approach for UEP by partial superposition transmission~(referred to as UEP-by-PST for convenience) using LDPC codes. In the UEP-by-PST system, the information sequence is distinguished as two parts, the MID and the LID, both of which are coded with binary LDPC codes. The codeword that corresponds to the MID is superimposed on the codeword that corresponds to the LID. The transmitted sequence consists of two parts. One is the codeword that corresponds to the MID, and the other is the superposition of the respective codewords that correspond to the MID and the LID. We then propose a decoding algorithm of the UEP-by-PST system, which can be described as an iterative message processing/passing algorithm over a high level normal graph. Discretized density evolution is conducted to predict the convergence thresholds for the MID and the LID of the UEP-by-PST. Simulation results verify our analysis and show that, over both additive white Gaussian noise~(AWGN) channels and uncorrelated Rayleigh fading channels, UEP-by-PST provides higher coding gain for the MID compared with the traditional EEP approach, but with negligible performance loss for the LID. To compare the UEP-by-PST with other approaches, we propose to use as a criterion the minimum signal-to-noise ratio~(SNR) that is required to guarantee the qualities of both the MID and the LID. Simulation results show that, under this practical criterion, UEP-by-PST provides more efficient UEP compared with the UEP approach in the digital video broadcasting~(DVB) system~\cite{DVB04}, which is referred to as UEP-by-Mapping in this paper.



The rest of this paper is organized as follows. We present the encoding and decoding algorithms of the UEP-by-PST system in Section~\ref{sec:UEP-by-PST}. Also given in Section~\ref{sec:UEP-by-PST} is the algebraic structure of the UEP-by-PST. In Section~\ref{sec:Analysis}, we present the asymptotic performance analysis of the UEP-by-PST. Numerical results are provided in Section~\ref{sec:Results}. Section~\ref{sec:Conclusion} concludes this paper.

\section{Unequal Error Protection by Partial Superposition Transmission}\label{sec:UEP-by-PST}
\subsection{Encoding Algorithm}
\begin{figure}
  \center
  \includegraphics[width=\figwidth]{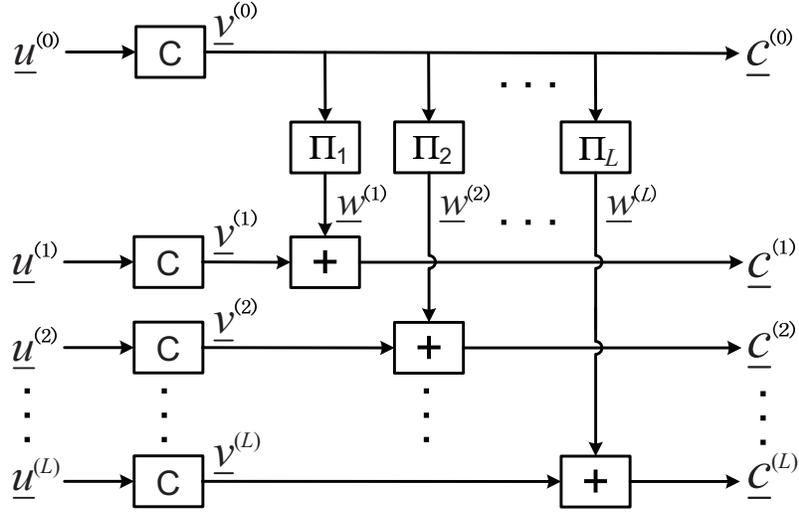}
  \caption{Encoding structure of the UEP-by-PST system.}
  \label{fig:encoder}
\end{figure}

Consider a binary LDPC code $\mathscr{C}[n, k]$ with dimension $k$ and length $n$, which is referred to as the {\em basic code} in this paper for convenience. Assume that the information sequence ${\underline u}$ can be equally grouped into $L+1$ blocks,
\begin{equation}\label{1-1}
{\underline u} = ({\underline u^{(0)}}, {\underline u^{(1)}}, \cdots, {\underline u^{(L)}}),
\end{equation}
where ${\underline u^{(0)}}$ and $({\underline u^{(1)}}, \ldots, {\underline u^{(L)}})$ are the MID of length $k$ and the LID of length $kL$, respectively. The encoding algorithm of the UEP-by-PST is described as follows, see Fig.~\ref{fig:encoder} for reference.

\vspace{0.1cm}
\begin{algorithm}{Encoding of the UEP-by-PST System}\label{Algorithm1}
\begin{itemize}
    \item {\bf{Encoding}:} For $0\leq \ell \leq L$, encode $\underline{u}^{(\ell)}$ into $\underline{v}^{(\ell)} \in \mathbb{F}_2^n$ by the~(systematic) encoding algorithm of the basic code $\mathscr{C}$.

    \item {\bf{Interleaving}:} For $1\leq \ell \leq L$, interleave $\underline{v}^{(0)}$ by the $\ell$-th interleaver $\mathbf{\Pi}_{\ell}$ of size $n$ into $\underline{w}^{(\ell)}$.

    \item {\bf{Superposition}:} For $1\leq \ell \leq L$, compute $\underline{c}^{(\ell)} = \underline{w}^{(\ell)} \oplus \underline{v}^{(\ell)}$, where $``\oplus "$ represents component-wise modulo-2 addition.

    \item {\bf{Combining}:} Output sequence $\underline{c} = (\underline{c}^{(0)}, \underline{c}^{(1)}, \cdots, \underline{c}^{(L)})$ of length $N$, where ${\underline c^{(0)}} = {\underline v^{(0)}}$ and $N=n(L+1)$.
\end{itemize}
\end{algorithm}

\textbf{Remarks:}
\begin{itemize}
  \item In principle, the basic code $\mathscr{C}$ can be chosen as any other types of codes, such as convolutional codes and turbo-like codes.
  \item The basic code $\mathscr{C}$ can also be chosen as a UEP code. In this case, the proposed UEP-by-PST system provides multilevel UEP.
\end{itemize}

\subsection{Algebraic Structure}\label{sec:AlgebraStructure}
Let $\mathbf{G}$ and $\mathbf{H}$ be the generator matrix and the parity-check matrix of the basic code $\mathscr{C}$, respectively. Let $\mathbf{\Pi}_{\ell}(\ell=1, \cdots, L)$ be a permutation matrix of size $n\times n$ corresponding to the $\ell$-th interleaver in Fig.~\ref{fig:encoder}. The encoding process of the UEP-by-PST system can be expressed as
\begin{eqnarray}\label{eq:G}
    \underline{c}^{(\ell)} &=& \left\{
            \begin{array}{ll}
               \underline{v}^{(0)}, & \ell = 0\\
               \underline{v}^{(0)}\mathbf{\Pi}_{\ell} \oplus \underline{v}^{(\ell)}, & 1 \leq \ell \leq L
            \end{array}\right.  \nonumber\\
            &=&\left\{
            \begin{array}{ll}
               \underline{u}^{(0)}\mathbf{G}, & \ell = 0\\
               \underline{u}^{(0)}\mathbf{G}\mathbf{\Pi}_{\ell} \oplus \underline{u}^{(\ell)}\mathbf{G}, & 1 \leq \ell \leq L
            \end{array}.\right.
\end{eqnarray}
Rewriting (\ref{eq:G}), we can get
\begin{eqnarray}\label{eq:G2}
\underline{c} &=& \left( \underline{c}^{(0)}, \underline{c}^{(1)}, \cdots, \underline{c}^{(L)} \right) \nonumber \\
             &=& \left( \underline{u}^{(0)}, \underline{u}^{(1)}, \cdots, \underline{u}^{(L)} \right) \cdot
             \mathbf{G}_{\text{\tiny \rm UEP-by-PST}},
\end{eqnarray}
where
\begin{eqnarray}\label{eq:G3}
\mathbf{G}_{\text{\tiny \rm UEP-by-PST}} =
    \left[\begin{array}{cccc}
        \mathbf{G}  &\mathbf{G}\mathbf{\Pi}_{1} &\cdots &\mathbf{G}\mathbf{\Pi}_{L}\\
            &\mathbf{G} &  &\\
            &           &\ddots &\\
            &           & &\mathbf{G}
        \end{array}
    \right]
\end{eqnarray}
is the generator matrix of the UEP-by-PST system. Let $\mathbf{H}_{\text{\tiny \rm UEP-by-PST}}$ be the parity-check matrix of the UEP-by-PST system. Since
\begin{eqnarray}
    \mathbf{G}_{\text{\tiny \rm UEP-by-PST}} \cdot \mathbf{H}_{\text{\tiny \rm UEP-by-PST}}^{\rm{T}} = \mathbf{0},
\end{eqnarray}
the parity-check matrix $\mathbf{H}_{\text{\tiny \rm UEP-by-PST}}$ can be represented as
\begin{eqnarray}\label{eq:H}
\mathbf{H}_{\text{\tiny \rm UEP-by-PST}} =
    \left[\begin{array}{cccc}
        \mathbf{H}  & & &\\
        \mathbf{H}\mathbf{\Pi}_{1}    &\mathbf{H} &  &\\
           \vdots &           &\ddots &\\
        \mathbf{H}\mathbf{\Pi}_{L}    &           & &\mathbf{H}
        \end{array}
    \right].
\end{eqnarray}

\subsection{Normal Graphical Realizations}\label{sec:NormalGraphNotation}

    \begin{figure}
        \center
        \includegraphics[width=\figwidth]{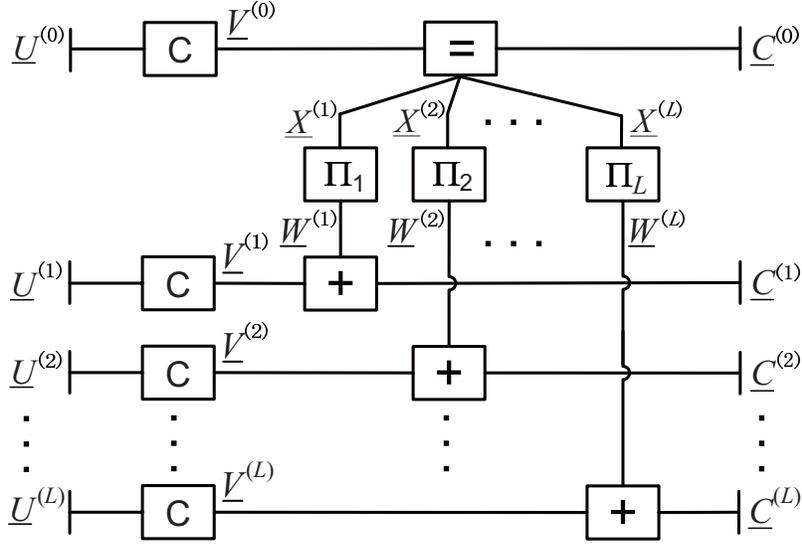}
        \caption{Normal realization of the UEP-by-PST system.}
        \label{fig:decoder}
    \end{figure}

The proposed UEP-by-PST system can be represented by a high-level normal graph~\cite{Forney01,Ma04}. In a general normal graph, {\em edges} represent {\em variables}, while {\em vertices} represent {\em constraints}. As shown in Fig.~\ref{fig:decoder}, there are four types of nodes in the normal graph of the UEP-by-PST system.

\begin{itemize}
  \item{\emph{Node} \fbox{C}:} The node \fbox{C} represents the constraint that ${\underline V^{(\ell)}}$ must be a codeword of $\mathscr{C}$ that corresponds to ${\underline U^{(\ell)}}$, for $0\leq \ell \leq L$. In practice, ${\underline U^{(\ell)}}$ is usually assumed to be independent and uniformly distributed over $\mathbb{F}_2^k$. Assume that the messages associated with ${\underline V^{(\ell)}}$ are available from the node \fbox{=} (when $\ell=0$) or the node \fbox{+} (when $1\leq \ell \leq L$). The node \fbox{C} performs the sum-product algorithm (SPA)~\cite{Kschischang01} to compute the extrinsic messages. The extrinsic messages associated with ${\underline V^{(\ell)}}$ are fed back to the node \fbox{=} (when $\ell=0$) or the node \fbox{+} (when $1\leq \ell \leq L$), while the extrinsic messages associated with ${\underline U^{(\ell)}}$ can be used to make decisions on the transmitted data.

  \item{\emph{Node} \fbox{=}:} The node \fbox{=} represents the constraint that all connecting variables must take the same realizations. The message processing/passing algorithm of the node \fbox{=} is the same as that of the variable node in a binary LDPC code.

  \item{\emph{Node} \fbox{$\Pi_{\ell}$}:} The node \fbox{$\Pi_{\ell}$} represents the $\ell$-th interleaver, which interleaves or de-interleaves the input messages.

  \item{\emph{Node} \fbox{+}:} The node \fbox{+} represents the constraint that all connecting variables must be added up to zero over $\mathbb{F}_2$. The message processing/passing algorithm of the node \fbox{+} is similar to that of the check node in a binary LDPC code. The only difference is that the messages associated with the half edge are available from the channel observations.
\end{itemize}

Then the normal graphical realization of the UEP-by-PST system can be divided into $L+1$ {\em layers}, one MID layer and $L$ LID layers, where the MID layer consists of a node of type \fbox{C} and a node of type \fbox{=}, while each LID layer consists of a node of type \fbox{C}, a node of type \fbox{+} and a node of type \fbox{$\Pi$}, see Fig.~\ref{fig:decoder} for reference.

\subsection{Decoding Algorithm}

A {\em message} associated with a discrete variable is defined as its probability mass function~(pmf) here. We focus on random variables defined over $\mathbb{F}_2$. For example, a message associated with a random variable $X$ over $\mathbb{F}_2$ can be represented by a real vector $P_X(x), x\in \mathbb{F}_2$, such that $P_X(0)+P_X(1)=1$. Let $X$ be a random variable corresponding to the edge connecting two vertices $\mathcal{A}$ and $\mathcal{B}$. We use the notation $P_{X}^{(\mathcal{A} \rightarrow \mathcal{B})}(x), x\in \mathbb{F}_2$~\cite{Ma12} to indicate the direction of the message flow.

\begin{figure}
    \centering
    \includegraphics[width=\figwidth]{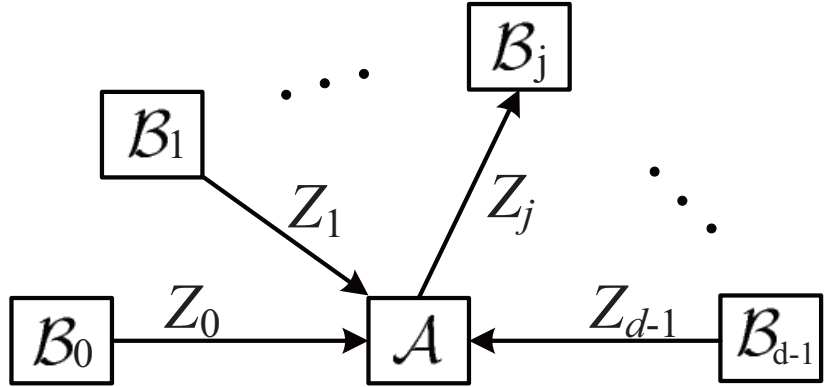}
    \caption{A generic node $\mathcal{A}$ as a message processor.}
\label{normalgraph}
\end{figure}

To describe the algorithm more clearly, we introduce a basic rule for message processing at an arbitrary node. Let $\mathcal{A}$ be a node connecting to $ \mathcal{B}_j$ with random variables $Z_j$ defined over $\mathbb{F}_2$~($0\leq j \leq d-1$), as shown in Fig.~\ref{normalgraph}. Assume that all incoming messages are available, which are denoted by $P_{Z_j}^{(\mathcal{B}_j\rightarrow \mathcal{A})}(z), z\in \mathbb{F}_2$. The node $\mathcal{A}$, as a {\em message processor}, delivers the outgoing message with respect to any given $Z_j$ by computing the likelihood function
    \begin{equation}\label{likelihood-function}
            P_{Z_j}^{(\mathcal{A}\rightarrow \mathcal{B}_j)}(z) \propto
            {\rm Pr}\{{ \mathcal{A}\; {\rm is\; satisfied}} \mid Z_j =
            z\}, \;\;\; z \in \mathbb{F}_2.
    \end{equation}
Because the computation of the likelihood function is irrelevant to the incoming message $P_{Z_j}^{(\mathcal{B}_j\rightarrow
\mathcal{A})}(z)$, we claim that $P_{Z_j}^{( \mathcal{A}\rightarrow\mathcal{B}_j)}(z)$ is exactly the so-called {\em extrinsic message}.

For simplicity, we assume that the codeword ${\underline{c}}$ of length $N$ is modulated and transmitted over a discrete memoryless channel, resulting in a received vector $\underline{y}$. In more general settings, we assume that the {\it a posteriori} probabilities ${\rm Pr}\{C_i = 0, 1 | \underline{y}\}$, $0 \leq i < N$ are computable\footnote{The computation in this step is irrelevant to the code constraints but depends only on the modulation and the channel.}, where $C_i$ is the $i$-th component of $\underline{C}$. Then, these {\it a posteriori} probabilities are used to initialize the decoding algorithm of the UEP-by-PST,
\begin{equation}\label{Init0}
    P_{C_j^{(0)}}^{\left( | \rightarrow = \right)}\left( c_j \right) = {\rm Pr}\{C_j^{(0)}=c_j | \underline{y}\},~c_j \in \mathbb{F}_2
\end{equation}
for $0 \leq j \leq n-1$, and
\begin{equation}\label{Init1}
    P_{C_j^{(\ell)}}^{\left( | \rightarrow + \right)}\left( c_{n\ell+j} \right) = {\rm Pr}\{C_j^{(\ell)}=c_{n\ell+j} | \underline{y}\},~c_{n\ell+j} \in \mathbb{F}_2
\end{equation}
for $0 \leq j \leq n-1$ and $1 \leq \ell \leq L$.

The iterative decoding algorithm of the UEP-by-PST can be described as an iterative message processing/passing algorithm over a high-level normal graph scheduled as follows, see Fig.~\ref{fig:decoder} for reference.

\vspace{0.15cm}
\begin{algorithm}{Iterative Decoding of the UEP-by-PST System}\label{Algorithm2}
\begin{itemize}
    \item {\bf{Initialization}:} All messages over the intermediate edges are initialized as uniformly distributed variables. Initialize the messages $P_{\underline{C}^{(0)}}^{\left( | \rightarrow = \right)}\left( \underline{c}^{(0)} \right)$ and $P_{\underline{C}^{(\ell)}}^{\left( | \rightarrow + \right)}\left( \underline{c}^{(\ell)} \right)$ for $1 \leq \ell \leq L$ according to~(\ref{Init0}) and~(\ref{Init1}), respectively. Select a maximum local iteration number $I_{max}>0$ and a maximum global iteration number $J_{max}>0$. Set $J=0$.

    \item {\bf{Iteration}:} While $J<J_{max}$
    \begin{enumerate}
        \item The MID layer performs a message processing/passing algorithm scheduled as
    \begin{equation*}
      \begin{array}{l}
        \fbox{=} \rightarrow \fbox{$\rm{C}$} \rightarrow \fbox{=}.
      \end{array}
    \end{equation*}
    To be more specific, at node $\fbox{=}$, compute the extrinsic messages $P_{\underline{V}^{(0)}}^{\left( = \rightarrow \rm{C} \right)}\left( \underline{v}^{(0)} \right)$; at node $\fbox{\rm{C}}$, perform the SPA for the basic code $\mathscr{C}$ with maximum local iteration number $I_{max}$ and compute the extrinsic messages $P_{\underline{V}^{(0)}}^{\left( \rm{C} \rightarrow = \right)}\left( \underline{v}^{(0)} \right)$; at node $\fbox{=}$, compute the
    extrinsic messages $P_{\underline{X}^{(\ell)}}^{\left( = \rightarrow \Pi_{\ell} \right)}\left( \underline{x}^{(\ell)} \right)$, for $1\leq \ell \leq L$.

        \item For $1\leq \ell \leq L$, the $\ell$-th LID layer performs a message processing/passing algorithm scheduled as
    \begin{equation*}
          \begin{array}{l}
            \fbox{$\Pi_{\ell}$} \rightarrow \fbox{+} \rightarrow
            \fbox{$\rm{C}$} \rightarrow \fbox{+} \rightarrow \fbox{$\Pi_{\ell}$}.
          \end{array}
    \end{equation*}
    To be more specific, at node \fbox{$\Pi_{\ell}$}, interleave the messages $P_{\underline{X}^{(\ell)}}^{\left( = \rightarrow \Pi_{\ell} \right)}\left( \underline{x}^{(\ell)} \right)$ into the messages $P_{\underline{W}^{(\ell)}}^{\left( \Pi_{\ell} \rightarrow + \right)}\left( \underline{w}^{(\ell)} \right)$; at node \fbox{+}, compute the extrinsic messages $P_{\underline{V}^{(\ell)}}^{\left( + \rightarrow \rm{C} \right)}\left( \underline{v}^{(\ell)} \right)$; at node $\fbox{\rm{C}}$, perform the SPA for the basic code $\mathscr{C}$ with maximum local iteration number $I_{max}$ and compute the extrinsic messages $P_{\underline{V}^{(\ell)}}^{\left( \rm{C} \rightarrow + \right)}\left( \underline{v}^{(\ell)} \right)$; at node \fbox{+}, compute the extrinsic messages $P_{\underline{W}^{(\ell)}}^{\left( + \rightarrow \Pi_{\ell} \right)}\left( \underline{w}^{(\ell)} \right)$; at node \fbox{$\Pi_{\ell}$}, deinterleave the messages $P_{\underline{W}^{(\ell)}}^{\left( + \rightarrow \Pi_{\ell} \right)}\left( \underline{w}^{(\ell)} \right)$ into the messages $P_{\underline{X}^{(\ell)}}^{\left( \Pi_{\ell} \rightarrow = \right)}\left( \underline{x}^{(\ell)} \right)$.

        \item For $0\leq \ell \leq L$, compute the full messages $P_{\underline{V}^{(\ell)}}\left( \underline{v}^{(\ell)} \right)$ as
    \begin{eqnarray}
    \begin{array}{l}
         P_{\underline{V}^{(\ell)}}\left( \underline{v}^{(\ell)} \right) \propto 
         \left\{
            \begin{array}{ll}
               P_{\underline{V}^{(0)}}^{\left( \rm{C} \rightarrow = \right)}\left( \underline{v}^{(0)} \right)P_{\underline{V}^{(0)}}^{\left( = \rightarrow \rm{C} \right)}\left( \underline{v}^{(0)} \right), & \ell = 0\\
               P_{\underline{V}^{(\ell)}}^{\left( \rm{C} \rightarrow + \right)}\left( \underline{v}^{(\ell)} \right)P_{\underline{V}^{(\ell)}}^{\left( + \rightarrow \rm{C} \right)}\left( \underline{v}^{(\ell)} \right), & 1 \leq \ell \leq L
            \end{array}\right.;
            \end{array}
    \end{eqnarray}
        then make hard decisions on ${\underline v^{(\ell)}}$ resulting in ${\underline {\hat{v}}^{(\ell)}}$; if all ${\underline {\hat{v}}^{(\ell)}}$ are valid codewords, declare the decoding successful, output ${\underline {\hat{u}}^{(\ell)}}$ for $0\leq \ell \leq L$, and exit the iteration.

        \item Increment $J$ by one.
    \end{enumerate}

    \item {\bf{Failure Report}:} If $J=J_{max}$, output ${\underline {\hat{u}}^{(\ell)}}$ for $0\leq \ell \leq L$ and report a decoding failure.
\end{itemize}
\end{algorithm}

\section{Asymptotic Performance Analysis}\label{sec:Analysis}
Density evolution, which was developed by Richardson and Urbanke~\cite{Richardson01}, is an effective analysis tool for computing the noise tolerance thresholds and optimizing degree sequences~\cite{Richardson01_1} of LDPC codes. In this section, discretized density evolution~\cite{Chung01} is conducted to predict the convergence thresholds for the MID and the LID of the UEP-by-PST.

Assume that all-zero codeword is transmitted over the AWGN channel with binary phase-shift keying~(BPSK) modulation and noise variance $\sigma^2$. To describe the density evolution, it is convenient to represent the message as in its equivalent form, the so-called {\em log-likelihood ratio}~(LLR). For example, the message computed in~(\ref{likelihood-function}) can be denoted as
\begin{equation}
    L_{Z_j}^{(\mathcal{A}\rightarrow \mathcal{B}_j)} \stackrel{\Delta}{=} \log \left( \frac{P_{Z_j}^{(\mathcal{A}\rightarrow \mathcal{B}_j)}(0)}{P_{Z_j}^{(\mathcal{A}\rightarrow \mathcal{B}_j)}(1)}\right).
\end{equation}
The LLR messages from the channel can be computed as~\cite{Moon05}
\begin{eqnarray}
  L\left(C\right) &\stackrel{\Delta}{=}& \log \left( \frac{{\rm Pr}\{C = 0 | y\}}{{\rm Pr}\{C = 1 |y\}} \right)\nonumber\\
  &=& \frac{2}{\sigma^2}y.
\end{eqnarray}
Let $Q(x)$ be the quantized message of $x$, i.e.,
\begin{equation}\label{quantization}
    Q( x ) \stackrel{\Delta}{=} \left\{\begin{array}{ll}
            -(2^{b-1} -1)\cdot\Delta, & \frac{x}{\Delta} \leq -(2^{b-1} -1)\\
            \left[\frac{x}{\Delta}\right]\cdot\Delta, & -(2^{b-1} -1) < \frac{x}{\Delta} < 2^{b-1} -1\\
            (2^{b-1} -1)\cdot\Delta, & \frac{x}{\Delta} \geq 2^{b-1} -1
            \end{array}\right. ,
\end{equation}
where $Q$ is the quantization operator, $b$ is the quantization bit, $\Delta$ is the quantization interval, and $[w]$ denotes the nearest integer to the real $w$.


For convenience, we define two sets $\mathcal{Q}=\{ i\cdot\Delta: -(2^{b-1} -1) \leq i \leq 2^{b-1} -1 \}$ and $\mathcal{L}=\{ \ell: 1 \leq \ell \leq L \}$. Assume that the interleavers $\mathbf{\Pi}_{\ell}$ are very large and random. With this assumption, we can investigate the ensemble of the UEP-by-PST system.

\begin{itemize}
  \item At node $\fbox{=}$, the message updating rule from node $\fbox{=}$ of degree $L+2$ to node \fbox{C} can be simply written as
\begin{equation}
    L_{\underline{V}^{(0)}}^{\left( = \rightarrow \rm{C} \right)} = L_{\underline{C}^{(0)}}^{\left( | \rightarrow = \right)} + \sum_{\ell \in \mathcal{L}} L_{\underline{X}^{(\ell)}}^{\left( + \rightarrow = \right)}.
\end{equation}
The messages $L_{\underline{C}^{(0)}}^{\left( | \rightarrow = \right)}$ are assumed to be identical independent distributed~(i.i.d.) variables with initial pmf
\begin{equation}
    P_{L_{C^{(0)}}^{\left( | \rightarrow = \right)}}\left( r \right) = {\rm Pr}\left\{Q\left( L\left(C^{(0)} \right)\right)=r \right\},
\end{equation}
while the messages $L_{\underline{X}^{(\ell)}}^{\left( + \rightarrow = \right)}$ for $1\leq \ell \leq L$ are assumed to be i.i.d. variables with initial pmf
\begin{eqnarray}
         P_{L_{X^{(\ell)}}^{\left( + \rightarrow = \right)}}\left( r \right) =
         \left\{
            \begin{array}{ll}
               1, & r = 0\\
               0, & r \neq 0
            \end{array}\right..
\end{eqnarray}
Thus, the pmf of $L_{\underline{V}^{(0)}}^{\left( = \rightarrow \rm{C} \right)}$ can be
determined as
\begin{equation}
 P_{L_{V^{(0)}}^{\left( = \rightarrow \rm{C} \right)}} = \mathcal{S}\left(P_{L_{C^{(0)}}^{\left( | \rightarrow = \right)}}, P_{\sum\limits_{\ell \in \mathcal{L}}L_{X^{(\ell)}}^{\left( + \rightarrow = \right)}}\right),
\end{equation}
where, for any two given pmfs $P_X$ and $P_{X'}$, the
transformation $\mathcal{S}$ is defined as
\begin{equation}
\mathcal{S}(P_X, P_{X'})(t) =  \sum_{(x,x'): t = Q(x+x')}
P_X(x)P_{X'}(x'),
\end{equation}
with $x,x', t\in \mathcal{Q}$. Since the messages $L_{\underline{X}^{(\ell)}}^{\left( + \rightarrow = \right)}$ for $1\leq \ell \leq L$ are i.i.d., the identical pmf is simply denoted by $P_{L_{X}^{\left( + \rightarrow = \right)}}$. Hence, the pmf $P_{\sum\limits_{\ell \in \mathcal{L}}L_{X^{(\ell)}}^{\left( + \rightarrow = \right)}}$ can be determined recursively as
\begin{eqnarray}
    P_{\sum\limits_{\ell \in \mathcal{L}}L_{X^{(\ell)}}^{\left( + \rightarrow = \right)}}
    &\stackrel{\Delta}{=}& \mathcal{S}^{L}P_{L_{X}^{\left( + \rightarrow = \right)}}\nonumber\\
    &=& \mathcal{S}\left(\mathcal{S}^{L-1}P_{L_{X}^{\left( + \rightarrow = \right)}},
    P_{L_{X}^{\left( + \rightarrow = \right)}}\right).~
\end{eqnarray}
Likewise, the message updating rule from node $\fbox{=}$ to node \fbox{+} can be simply written as
\begin{equation}
    L_{\underline{X}^{(\ell)}}^{\left( = \rightarrow + \right)} = L_{\underline{C}^{(0)}}^{\left( | \rightarrow = \right)} + L_{\underline{V}^{(0)}}^{\left( \rm{C} \rightarrow = \right)} + \sum_{\ell' \in \mathcal{L}\backslash \ell}L_{\underline{X}^{(\ell')}}^{\left( + \rightarrow = \right)} ,
\end{equation}
where the pmf of $L_{\underline{X}^{(\ell)}}^{\left( = \rightarrow + \right)}$ can be determined as
\begin{equation}
 P_{L_{X^{(\ell)}}^{\left( = \rightarrow + \right)}} = \mathcal{S}\left(P_{L_{C^{(0)}}^{\left( | \rightarrow = \right)} + L_{V^{(0)}}^{\left( \rm{C} \rightarrow = \right)}}, P_{\sum\limits_{\ell' \in \mathcal{L}\backslash \ell}L_{X^{(\ell')}}^{\left( + \rightarrow = \right)}}\right).
\end{equation}
The pmf $P_{L_{C^{(0)}}^{\left( | \rightarrow = \right)} + L_{V^{(0)}}^{\left( \rm{C} \rightarrow = \right)}}$ can be determined as
\begin{equation}
 P_{L_{C^{(0)}}^{\left( | \rightarrow = \right)} + L_{V^{(0)}}^{\left( \rm{C} \rightarrow = \right)}} = \mathcal{S}\left( P_{L_{C^{(0)}}^{\left( | \rightarrow = \right)}}, P_{L_{V^{(0)}}^{\left( \rm{C} \rightarrow = \right)}}\right),
\end{equation}
while the pmf $P_{\sum\limits_{\ell' \in \mathcal{L}\backslash \ell}L_{X^{(\ell')}}^{\left( + \rightarrow = \right)}}$ can be determined recursively as
\begin{eqnarray}
    P_{\sum\limits_{\ell' \in \mathcal{L}\backslash \ell}L_{X^{(\ell')}}^{\left( + \rightarrow = \right)}}
    &\stackrel{\Delta}{=}& \mathcal{S}^{L-1}P_{L_{X}^{\left( + \rightarrow = \right)}}\nonumber\\
    &=& \mathcal{S}\left(\mathcal{S}^{L-2}P_{L_{X}^{\left( + \rightarrow = \right)}},
    P_{L_{X}^{\left( + \rightarrow = \right)}}\right).
\end{eqnarray}

  \item At node $\fbox{+}$, the message updating rule from node $\fbox{+}$ of degree $3$ to node \fbox{C} can be simply written as
\begin{equation}
    L_{\underline{V}^{(\ell)}}^{\left( + \rightarrow \rm{C} \right)} = 
    2 \tanh^{-1} \left( \tanh \left( L_{\underline{C}^{(\ell)}}^{\left( | \rightarrow + \right)}/2 \right) \tanh \left(L_{\underline{X}^{(\ell)}}^{\left( = \rightarrow + \right)}/2 \right)\right).
\end{equation}
The messages $L_{\underline{C}^{(\ell)}}^{\left( | \rightarrow + \right)}$ are assumed to be i.i.d. variables with initial pmf
\begin{equation}
    P_{L_{C^{(\ell)}}^{\left( | \rightarrow + \right)}}\left( r \right) = {\rm Pr}\left\{Q\left( L\left(C^{(\ell)} \right)\right)=r\right\}.
\end{equation}
Thus, the pmf of $L_{\underline{V}^{(\ell)}}^{\left( + \rightarrow \rm{C} \right)}$ can be determined as
\begin{equation}\label{node+}
 P_{L_{V^{(\ell)}}^{\left( + \rightarrow \rm{C} \right)}} = \mathcal{T}\left(P_{L_{C^{(\ell)}}^{\left( | \rightarrow + \right)}}, P_{L_{X^{(\ell)}}^{\left( = \rightarrow + \right)}}\right),
\end{equation}
where, for any two given pmfs $P_X$ and $P_{X'}$, the
transformation $\mathcal{T}$ is defined as
\begin{eqnarray}
\mathcal{T}(P_X, P_{X'})(t) =
\sum\limits_{(x,x'): t = Q(2 \tanh^{-1} \left( \tanh \left( x/2 \right) \tanh \left( x'/2 \right)\right))} P_X(x)P_{X'}(x'),
\end{eqnarray}
with $x,x', t\in \mathcal{Q}$. Apparently, the pmfs $P_{L_{C^{(\ell)}}^{\left( | \rightarrow + \right)}}$ for $1\leq \ell \leq L$ are equal and hence denoted simply by $P_{L_{C}^{\left( | \rightarrow + \right)}}$. Similarly, we denote $P_{L_{X^{(\ell)}}^{\left( = \rightarrow + \right)}}$ and $P_{L_{V^{(\ell)}}^{\left( + \rightarrow \rm{C} \right)}}$ by $P_{L_{X}^{\left( = \rightarrow + \right)}}$ and $P_{L_{V}^{\left( + \rightarrow \rm{C} \right)}}$, respectively. Hence, the pmf $P_{L_{V^{(\ell)}}^{\left( + \rightarrow \rm{C} \right)}}$ can be computed as
\begin{eqnarray}
    P_{L_{V^{(\ell)}}^{\left( + \rightarrow \rm{C} \right)}} &\stackrel{\Delta}{=}& P_{L_{V}^{\left( + \rightarrow \rm{C} \right)}}\nonumber\\
    &=& \mathcal{T}\left(P_{L_{C}^{\left( | \rightarrow + \right)}}, P_{L_{X}^{\left( = \rightarrow + \right)}}\right).
\end{eqnarray}
Likewise, the message updating rule from node $\fbox{+}$ to node $\fbox{=}$ can be simply written as
\begin{eqnarray}
\begin{array}{l}
    L_{\underline{X}^{(\ell)}}^{\left( + \rightarrow = \right)} = 
    2 \tanh^{-1} \left( \tanh \left( L_{\underline{C}^{(\ell)}}^{\left( | \rightarrow + \right)}/2 \right) \tanh \left(L_{\underline{V}^{(\ell)}}^{\left( \rm{C} \rightarrow + \right)}/2 \right)\right),
\end{array}
\end{eqnarray}
where the pmf of $L_{X^{(\ell)}}^{\left( + \rightarrow = \right)}$ can be determined as
\begin{eqnarray}
    P_{L_{X^{(\ell)}}^{\left( + \rightarrow = \right)}} &\stackrel{\Delta}{=}& P_{L_{X}^{\left( + \rightarrow = \right)}}\nonumber\\
    &=& \mathcal{T}\left(P_{L_{C}^{\left( | \rightarrow + \right)}}, P_{L_{V}^{\left( \rm{C} \rightarrow + \right)}}\right).
\end{eqnarray}

  \item At node \fbox{C}, the message updating rule is the same as shown in~\cite{Chung01}. After a fixed number of
local iterations $I_{max}$, we can obtain the extrinsic messages $L_{\underline{V}^{(0)}}^{\left( \rm{C} \rightarrow = \right)}$, $L_{\underline{V}^{(\ell)}}^{\left( \rm{C} \rightarrow + \right)}$ for $1\leq \ell \leq L$ and their corresponding pmfs $P_{L_{V^{(0)}}^{\left( \rm{C} \rightarrow = \right)}}$, $P_{L_{V^{(\ell)}}^{\left( \rm{C} \rightarrow + \right)}}$ for $1\leq \ell \leq L$, respectively. We can also compute the full messages $L_{\underline{V}^{(\ell)}}$ and their corresponding pmfs $P_{L_{V^{(\ell)}}}$ for $0\leq \ell \leq L$.
\end{itemize}

In summary, for a given parameter $L$ and a local iteration number $I_{max}$, we can iteratively update the pmfs $P_{L_{\underline{V}^{(\ell)}}}$ for $0\leq \ell \leq L$ according to the decoding procedure scheduled as
    \begin{equation*}
        \fbox{=} \rightarrow \fbox{$\rm{C}$} \rightarrow \fbox{=} \rightarrow
        \fbox{+} \rightarrow \fbox{$\rm{C}$} \rightarrow \fbox{+} \rightarrow \fbox{=}.
    \end{equation*}
Therefore, we may determine~(by commonly-used one-dimensional search) the minimum $E_b/N_0$ such that the BER for the MID~(or the LID) tends to zero as the number of global iterations tends to infinity.

\section{Numerical Results}\label{sec:Results}
In this section, we first give the thresholds of the UEP-by-PST by using the discretized density evolution techniques. Then we compare the bit-error rate~(BER) performance of the UEP-by-PST with those of the traditional EEP approach over AWGN channels and uncorrelated Rayleigh fading channels. For the uncorrelated Rayleigh fading channels, we assume that the channel state information is available at the receiver. Finally, we compare the UEP-by-PST and the UEP-by-Mapping in the DVB system from a practical point of view.

\subsection{Thresholds of UEP-by-PST}


\begin{table*}
\caption{Thresholds of the UEP-by-PST Based on Density Evolution}\label{table1}
  \centering
  \begin{tabular}{|c||c|c|c|c|}
  \hline
  \multirow{2}{1.6cm}{Threshold $\rm{(E_b/N_0)}$} &\multicolumn{3}{c|}{UEP-by-PST} &\multirow{2}{*}{EEP}\\ \cline{2-4}
  & $L=1$ & $L=2$ & $L=3$ & \\ \hline \hline
  MID & 0.80~dB & 0.61~dB & 0.47~dB &\multirow{2}{*}{1.11~dB}\\ \cline{1-4}
  LID & 1.17~dB & 1.17~dB & 1.17~dB & \\
  \hline
\end{tabular}
\end{table*}

\begin{example}
Consider a random~$(3,6)$ regular LDPC code~\cite{Gallager63} with rate $1/2$ for the basic code $\mathscr{C}$. The local iteration number $I_{max}$ for LDPC decoding process is 50. The quantization interval $\Delta = 25/512$ with 10-bit quantization. Table~\ref{table1} gives the convergence thresholds for the MID and the LID of the UEP-by-PST. Also included in the table is the threshold of the traditional EEP approach. It can be seen that the thresholds for the LID with different $L$ ($1,~2$ and $3$) are the same. The gap of the thresholds for the MID between $L=1$ and $L=2$ is $0.19~\rm{dB}$, while that of the MID between $L=2$ and $L=3$ is $0.14~\rm{dB}$. From these thresholds, we can see that, UEP-by-PST theoretically provides higher coding gain for the MID compared with the traditional EEP approach, but with negligible performance loss for the LID.

\end{example}

\subsection{Performance of UEP-by-PST}
In the following examples, $L$ random interleavers, each of size $n$, are used for encoding. The iterative decoding algorithm of the UEP-by-PST is implemented with maximum global iteration number $J_{max}=20$ and maximum local iteration number $I_{max}=50$, while the iterative decoding algorithm of the traditional EEP approach is implemented with maximum iteration number $100$.

\begin{figure}
    \centering
    \includegraphics[width=\figwidth]{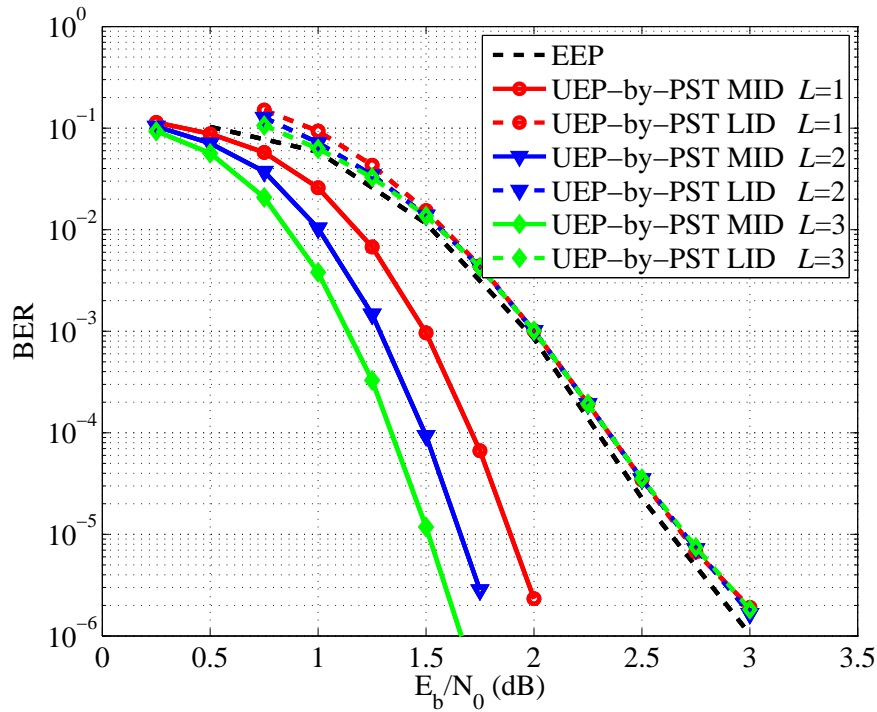}
    \caption{Performances of the UEP-by-PST with BPSK signalling over AWGN channels in Example 2. The basic code is a random~$(3,6)$ regular LDPC code with length $1024$.}
    \label{Fig3_1024_AWGN}
\end{figure}

\begin{figure}
    \centering
    \includegraphics[width=\figwidth]{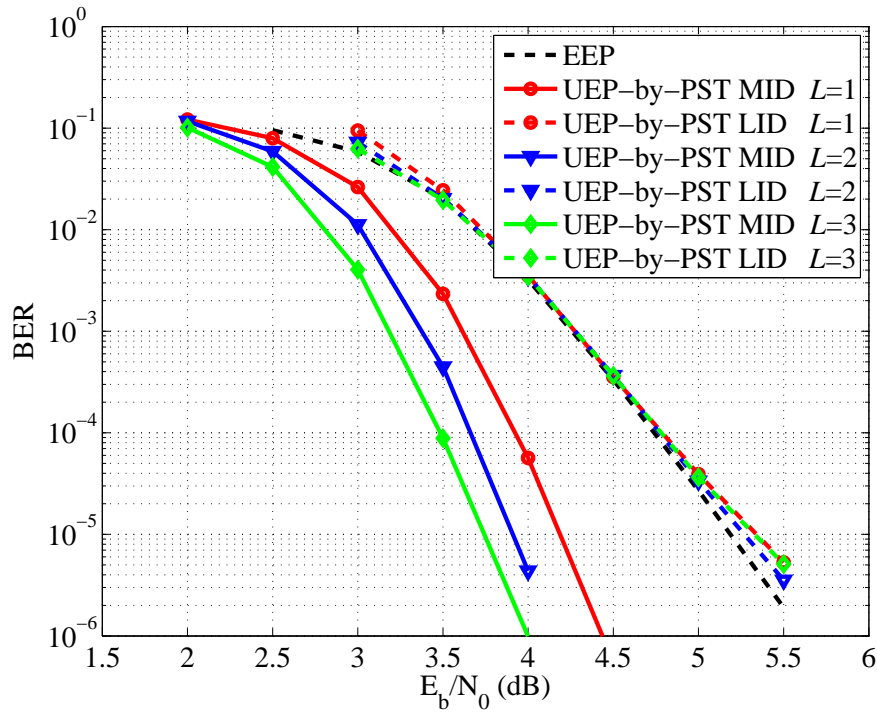}
    \caption{Performances of the UEP-by-PST with BPSK signalling over uncorrelated Rayleigh fading channels in Example 2. The basic code is a random~$(3,6)$ regular LDPC code with length $1024$.}
    \label{Fig3_1024_Rayleigh}
\end{figure}

\begin{example}
Consider a random~$(3,6)$ regular LDPC code~\cite{Gallager63} with length $1024$ for the basic code $\mathscr{C}$. The BER performances of the UEP-by-PST with BPSK signalling over AWGN channels and uncorrelated Rayleigh fading channels are shown in Fig.~\ref{Fig3_1024_AWGN} and Fig.~\ref{Fig3_1024_Rayleigh}, respectively. From Fig.~\ref{Fig3_1024_AWGN} and Fig.~\ref{Fig3_1024_Rayleigh}, we can see that, over both AWGN channels and uncorrelated Rayleigh fading channels, UEP-by-PST provides higher coding gain for the MID compared with the traditional EEP approach, but with negligible performance loss for the LID. For example, at $\rm{BER} = 10^{-5}$,
\begin{itemize}
  \item over AWGN channels, UEP-by-PST achieves about $0.7~\rm{dB},~1.0~\rm{dB}$ and $1.1~\rm{dB}$ extra coding gain for the MID compared with the traditional EEP approach when $L=1,~2$ and $3$, respectively;
  \item over uncorrelated Rayleigh fading channels, UEP-by-PST achieves about $1.0~\rm{dB},~1.3~\rm{dB}$ and $1.4~\rm{dB}$ extra coding gain for the MID compared with the traditional EEP approach when $L=1,~2$ and $3$, respectively.
\end{itemize}

\end{example}

\begin{figure}
    \centering
    \includegraphics[width=\figwidth]{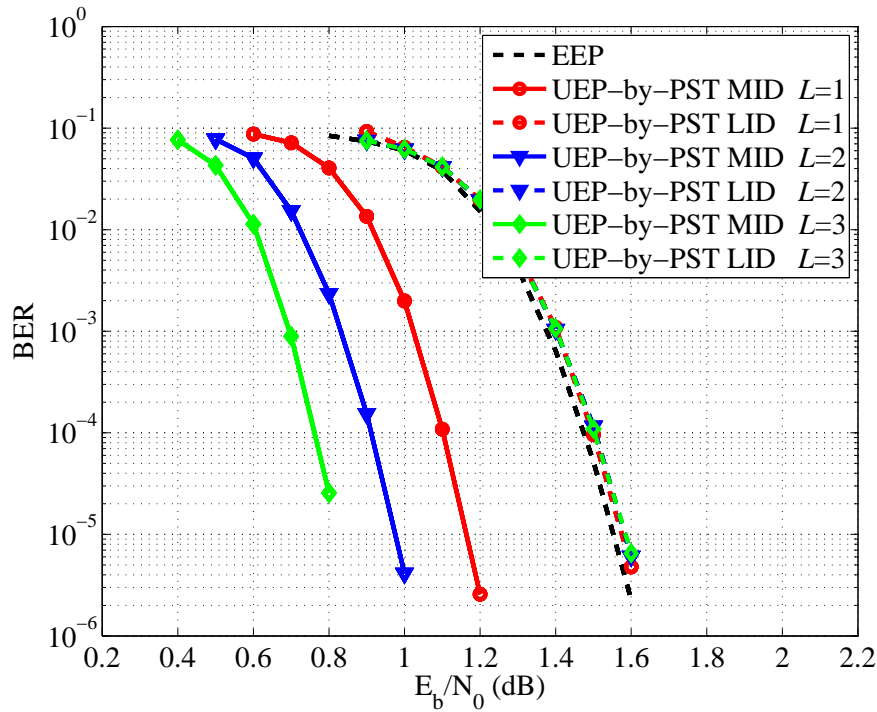}
    \caption{Performances of the UEP-by-PST with BPSK signalling over AWGN channels in Example 3. The basic code is a random~$(3,6)$ regular LDPC code with length $10000$.}
    \label{Fig4_10000_AWGN}
\end{figure}

\begin{figure}
    \centering
    \includegraphics[width=\figwidth]{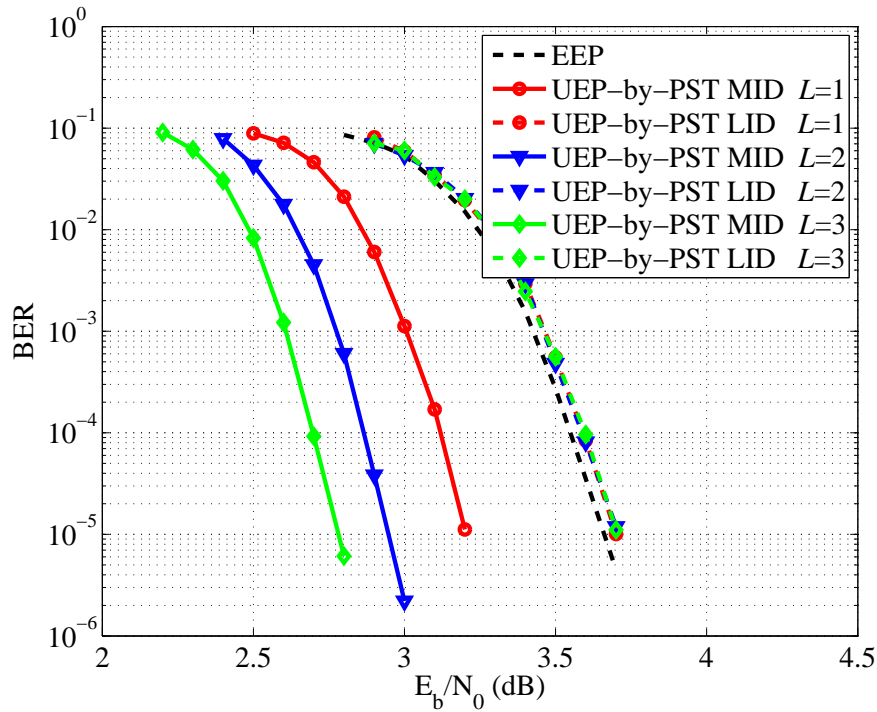}
    \caption{Performances of the UEP-by-PST with BPSK signalling over uncorrelated Rayleigh fading channels in Example 3. The basic code is a random~$(3,6)$ regular LDPC code with length $10000$.}
    \label{Fig4_10000_Rayleigh}
\end{figure}

\begin{example}
Consider a random~$(3,6)$ regular LDPC code with length $10000$ for the basic code $\mathscr{C}$. The BER performances of the UEP-by-PST with BPSK signalling over AWGN channels and uncorrelated Rayleigh fading channels are shown in Fig.~\ref{Fig4_10000_AWGN} and Fig.~\ref{Fig4_10000_Rayleigh}, respectively. From Fig.~\ref{Fig4_10000_AWGN} and Fig.~\ref{Fig4_10000_Rayleigh}, we can see that, over both AWGN channels and uncorrelated Rayleigh fading channels, UEP-by-PST provides higher coding gain for the MID compared with the traditional EEP approach, but with  negligible performance loss for the LID. For example, at $\rm{BER} = 10^{-4}$,
\begin{itemize}
  \item over AWGN channels, UEP-by-PST achieves about $0.4~\rm{dB},~0.6~\rm{dB}$ and $0.7~\rm{dB}$ extra coding gain for the MID compared with the traditional EEP approach when $L=1,~2$ and $3$, respectively;
  \item over uncorrelated Rayleigh fading channels, UEP-by-PST achieves about $0.4~\rm{dB},~0.7~\rm{dB}$ and $0.9~\rm{dB}$ extra coding gain for the MID compared with the traditional EEP approach when $L=1,~2$ and $3$, respectively.
\end{itemize}

\end{example}

\begin{figure}
    \centering
    \includegraphics[width=\figwidth]{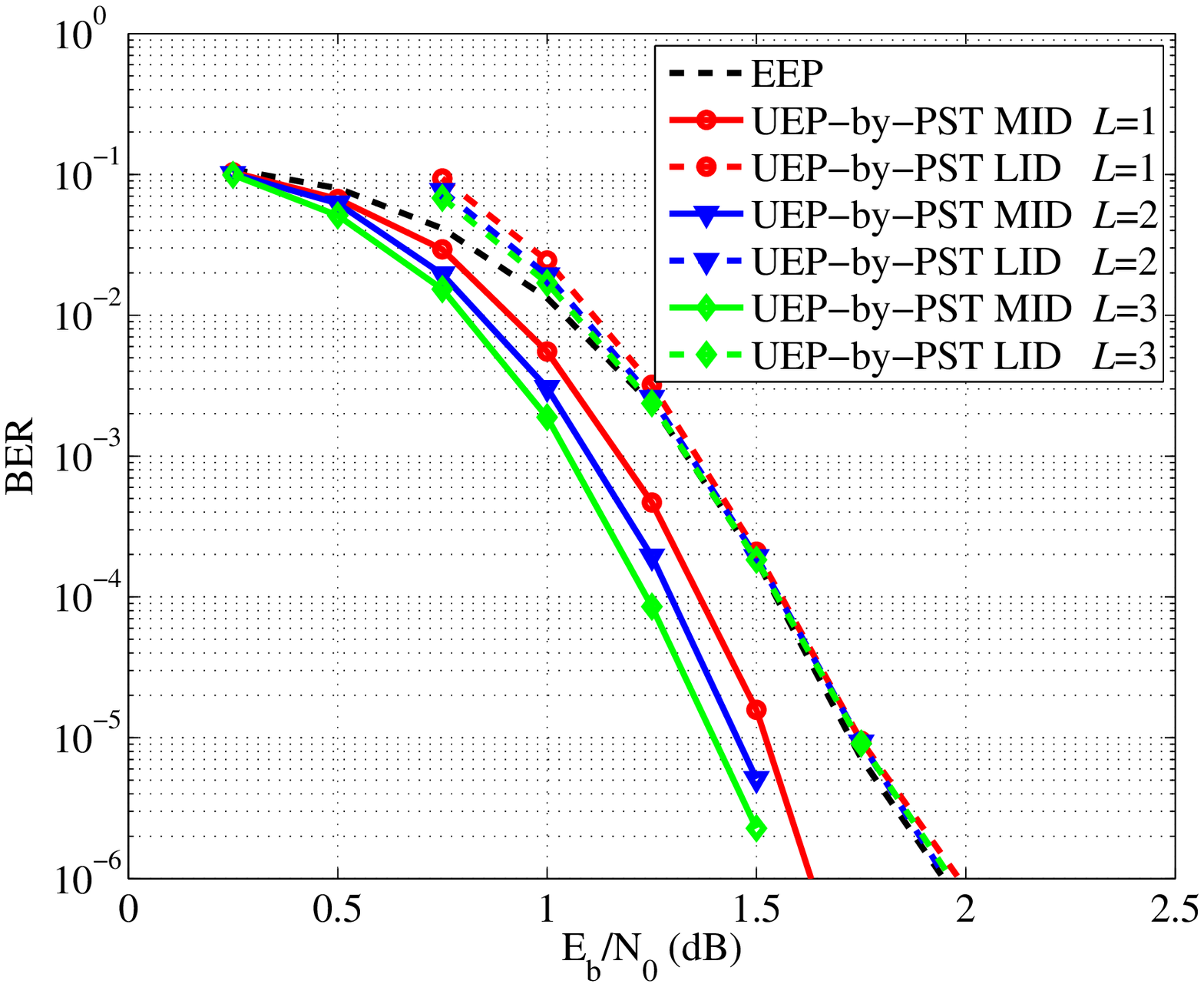}
    \caption{Performances of the UEP-by-PST with BPSK signalling over AWGN channels in Example 4. The basic code is an IEEE $802.11$n LDPC code with length $1944$ and rate $1/2$.}
    \label{Fig5_1944_AWGN}
\end{figure}

\begin{figure}
    \centering
    \includegraphics[width=\figwidth]{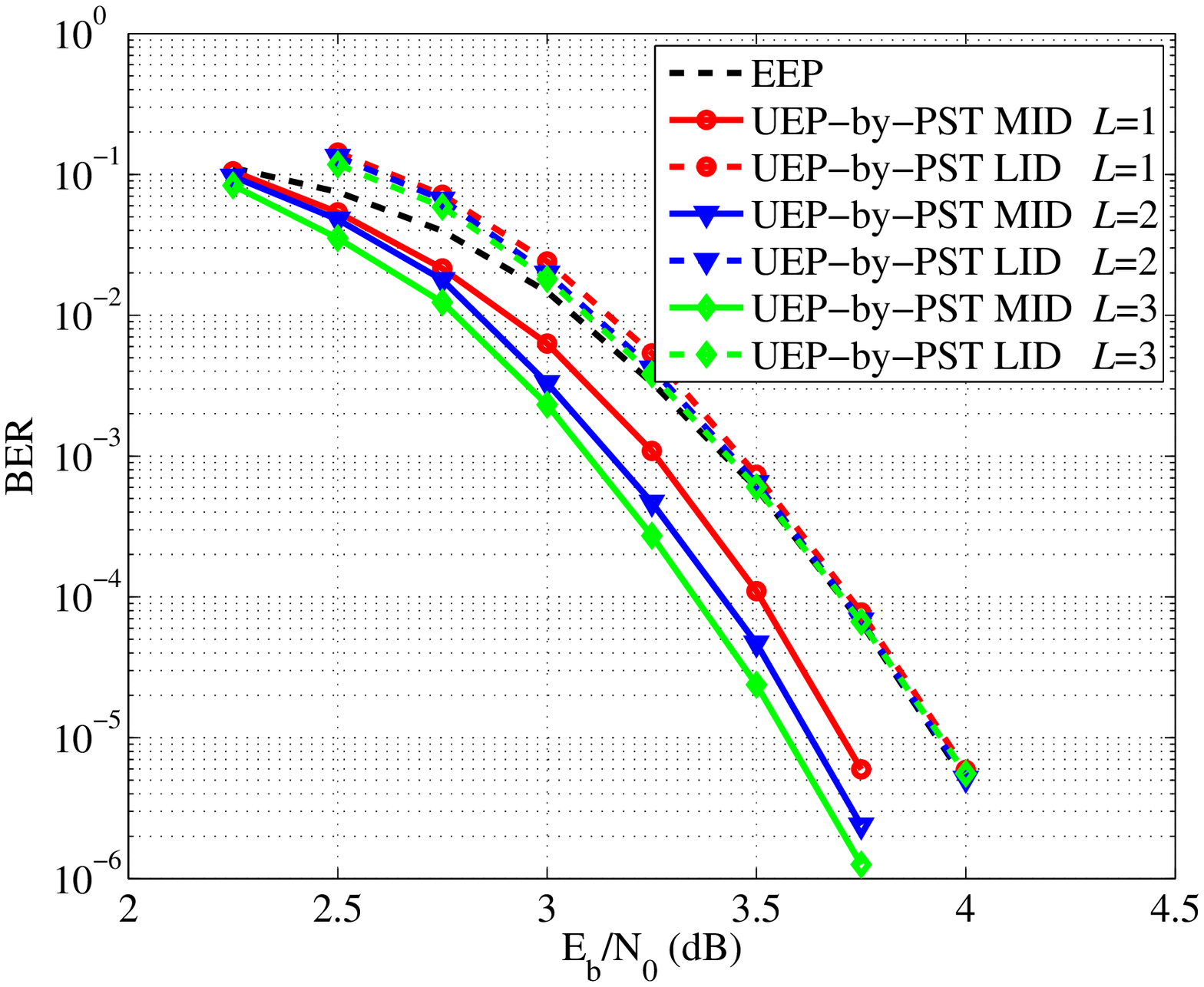}
    \caption{Performances of the UEP-by-PST with BPSK signalling over uncorrelated Rayleigh fading channels in Example 4. The basic code is an IEEE $802.11$n LDPC code with length $1944$ and rate $1/2$.}
    \label{Fig5_1944_Rayleigh}
\end{figure}

\begin{example}
Consider an IEEE $802.11$n LDPC code~\cite{IEEE80211n} with length $1944$ and rate $1/2$ for the basic code $\mathscr{C}$. The BER performances of the UEP-by-PST with BPSK signalling over AWGN channels and uncorrelated Rayleigh fading channels are shown in Fig.~\ref{Fig5_1944_AWGN} and Fig.~\ref{Fig5_1944_Rayleigh}, respectively. From Fig.~\ref{Fig5_1944_AWGN} and Fig.~\ref{Fig5_1944_Rayleigh}, we can see that, over both AWGN channels and uncorrelated Rayleigh fading channels, UEP-by-PST provides higher coding gain for the MID compared with the traditional EEP approach, but with negligible performance loss for the LID. For example, at $\rm{BER} = 10^{-5}$, over both AWGN channels and uncorrelated Rayleigh fading channels, UEP-by-PST achieves about $0.3~\rm{dB}$ extra coding gain for the MID compared with the traditional EEP approach, but with negligible performance loss for the LID.
\end{example}

\textbf{Remarks:}
\begin{itemize}
  \item From Fig.~\ref{Fig3_1024_AWGN} and Fig.~\ref{Fig4_10000_AWGN}, we can see that the extra coding gains at $\rm{BER} = 10^{-5}$ for the MID are similar to those at $\rm{BER} \rightarrow 0$ as predicted in Table~\ref{table1} by the discretized density evolution. We can also see that the performance loss for the LID is negligible, again as predicted by the discretized density evolution.

  \item Given the parity-check matrix~$\mathbf{H}_{\text{\tiny \rm UEP-by-PST}}$~(\ref{eq:H}), we can also perform directly the SPA over the corresponding normal graph. While, from our simulations, we have found that decoding the system as a single LDPC code specified by $\mathbf{H}_{\text{\tiny \rm UEP-by-PST}}$ delivers almost the same results.
\end{itemize}


\subsection{Comparison between UEP-by-PST and UEP-by-Mapping}
\begin{figure*}
    \centering
    \includegraphics[width=15cm]{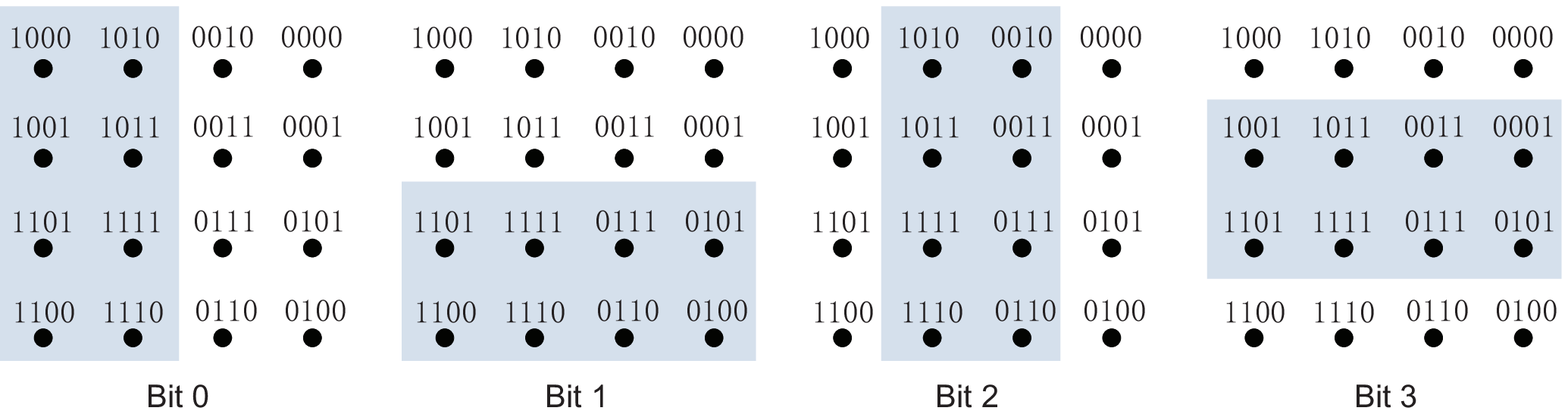}
    \caption{A 16-QAM mapping and the corresponding bit patterns.}
    \label{Fig-constellation}
\end{figure*}
In the following example, we consider a 16-QAM mapping scheme used in the DVB system~\cite{DVB04}. The mapping and its corresponding bit patterns are shown in Fig.~\ref{Fig-constellation}. In this mapping, if a signal whose label has the value ``0" in bit $m$, $0\leq m\leq 3$, then an error occurs if the received signal falls in the shaded region. As pointed out by Ayd{\i}nl{\i}k and Salehi~\cite{Aydinlik08}, in such a 16-QAM mapping scheme, the average number of nearest neighbors for bit 0, 1, 2 and 3 are 0.5, 0.5, 1.0 and 1.0, respectively. The first two bits are more protected than the last two bits. Apparently, this 16-QAM mapping can provide two levels of UEP.

Assume that the parameter $L=3$. In the UEP-by-PST, only the partial superposition transmission contributes to UEP. Bits of the codeword $\underline{c}^{(0)}$ and those of the codewords $(\underline{c}^{(1)}, \cdots, \underline{c}^{(L)})$ are transmitted in separate signaling intervals. That is, in the UEP-by-PST system, one 16-QAM signal point carries either four bits from the codeword $\underline{c}^{(0)}$ or four bits from the codewords $(\underline{c}^{(1)}, \cdots, \underline{c}^{(L)})$. In contrast, in the UEP-by-Mapping used in the DVB system~\cite{DVB04}, only the mapping contributes to UEP. That is, a bit of the codeword $\underline{c}^{(0)}$ and three bits of the codewords $(\underline{c}^{(1)}, \cdots, \underline{c}^{(L)})$ are mapped into one 16-QAM signal point, using the first bit position and the last three bit positions, respectively.

\begin{example}
Consider the same random~$(3,6)$ regular LDPC code with length $1024$ used in Example~2 for the basic code $\mathscr{C}$.
The BER performances of the UEP approaches~(UEP-by-PST and UEP-by-Mapping) with BPSK signalling over AWGN channels and uncorrelated Rayleigh fading channels are shown in Fig.~\ref{Fig_1024_16QAM_AWGN} and Fig.~\ref{Fig_1024_16QAM_Rayleigh}, respectively. The curve labeled ``EEP" shows the performance of the traditional EEP approach. From Fig.~\ref{Fig_1024_16QAM_AWGN} and Fig.~\ref{Fig_1024_16QAM_Rayleigh}, we can see that, over both AWGN channels and uncorrelated Rayleigh fading channels,
\begin{itemize}
  \item UEP-by-PST provides higher coding gain for the MID compared with the traditional EEP approach while causes negligible performance loss for the LID;
  \item UEP-by-Mapping provides higher coding gain for the MID compared with the traditional EEP approach but degrades the performance of the LID.
\end{itemize}
For example, at $\rm{BER} = 10^{-5}$, over both AWGN channels and uncorrelated Rayleigh fading channels,
\begin{itemize}
  \item UEP-by-PST achieves about $1.5~\rm{dB}$ extra coding gain for the MID compared with the traditional EEP approach, but with negligible performance loss for the LID;
  \item UEP-by-Mapping achieves about $2.4~\rm{dB}$ extra coding gain for the MID compared with the traditional EEP approach by sacrificing about $1.0~\rm{dB}$ coding gain for the LID.
\end{itemize}

\begin{figure}
    \centering
    \includegraphics[width=\figwidth]{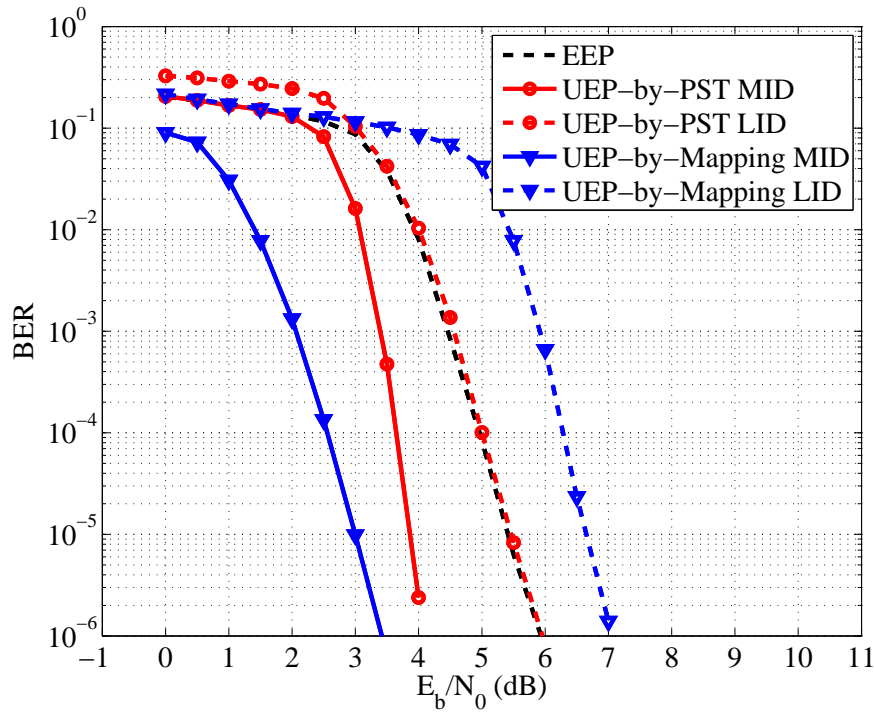}
    \caption{Simulation results of the UEP-by-PST and UEP-by-Mapping with 16-QAM over AWGN channels in Example 5. The basic code is a random~$(3,6)$ regular LDPC code with length $1024$. The parameter $L=3$.}
    \label{Fig_1024_16QAM_AWGN}
\end{figure}
\begin{figure}
    \centering
    \includegraphics[width=\figwidth]{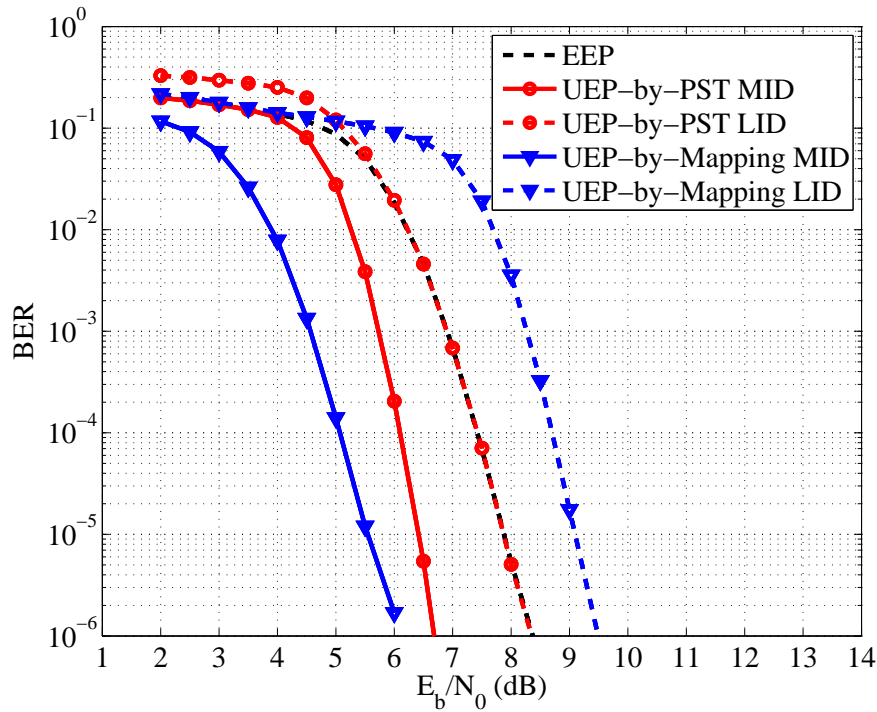}
    \caption{Simulation results of the UEP-by-PST and UEP-by-Mapping with 16-QAM over uncorrelated Rayleigh fading channels in Example 5. The basic code is a random~$(3,6)$ regular LDPC code with length $1024$. The parameter $L=3$.}
    \label{Fig_1024_16QAM_Rayleigh}
\end{figure}

From Fig.~\ref{Fig_1024_16QAM_AWGN} and Fig.~\ref{Fig_1024_16QAM_Rayleigh}, we can see that, UEP-by-PST is better than UEP-by-Mapping in terms of the LID, but worse than UEP-by-Mapping in terms of the MID. An interesting issue~(but rarely mentioned in the literatures) is how to compare different UEP approaches in terms of efficiency. To address this issue, we propose the following criterion from a practical perspective.

Assume that ($\varepsilon_0, \varepsilon_1$) are the error performance requirements by the MID and the LID, respectively. We denote the minimum SNR required for the MID and the LID by $\rm{SNR}(\varepsilon_0)$ and $\rm{SNR}(\varepsilon_1)$, respectively. Thus, the minimum SNR required for the UEP approach can be calculated as
\begin{equation}
    \rm{SNR}_{\text{\tiny \rm UEP}} = \max\left\{ \rm{SNR}(\varepsilon_0), \rm{SNR}(\varepsilon_1)\right\},
\end{equation}
which specifies the minimum SNR required to guarantee the qualities of both the MID and the LID. Hence, it can be taken as a criterion to compare different UEP approaches.


\begin{table*}
\caption{Minimum SNR's required by the UEP approaches}\label{table2}
  \centering
  \begin{tabular}{|c||c|c|c|}
  \hline
  \multirow{2}{*}{Minimum SNR} &\multirow{2}{*}{EEP} &\multirow{2}{*}{UEP-by-Mapping}  &\multirow{2}{*}{UEP-by-PST}\\
   & & & \\ \hline \hline
  AWGN          &5.4~dB   &4.9~dB            &3.9~dB\\ \hline
  Rayleigh      &7.9~dB   &7.0~dB            &6.4~dB\\ \hline
\end{tabular}
\end{table*}

We assume that $\varepsilon_0 \approx 10^{-5}$. From Fig.~\ref{Fig_1024_16QAM_AWGN} and Fig.~\ref{Fig_1024_16QAM_Rayleigh}, we can see that, over both AWGN channels and uncorrelated Rayleigh fading channels, $\rm{SNR}_{\text{\tiny \rm UEP-by-PST}} > \rm{SNR}_{\text{\tiny \rm UEP-by-Mapping}}$ when $\varepsilon_1 > 1.0 \times 10^{-1}$, while $\rm{SNR}_{\text{\tiny \rm UEP-by-PST}} < \rm{SNR}_{\text{\tiny \rm UEP-by-Mapping}}$ when $\varepsilon_1 < 1.0 \times 10^{-1}$. Suppose that we have an application that requires $\varepsilon_0 \approx 1.0 \times 10^{-5}$ and $\varepsilon_1 \approx 5.0 \times 10^{-2}$. Table~\ref{table2} gives the minimum SNRs required by the UEP approaches such that the error performance requirements ($\varepsilon_0, \varepsilon_1$) are simultaneously satisfied. Also included in the table is the minimum SNR required by the traditional EEP approach. From Table~\ref{table2}, we can see that, for these parameters,
\begin{itemize}
  \item UEP-by-PST performs 1.5~dB better than the traditional EEP approach over both AWGN channels and uncorrelated Rayleigh fading channels;
  \item UEP-by-PST performs 1.0~dB and 0.6~dB better than UEP-by-Mapping over AWGN channels and uncorrelated Rayleigh fading channels, respectively.
\end{itemize}
In summary, from a practical point of view, UEP-by-PST is an efficient approach to achieving UEP.

\end{example}

\section{Conclusion}\label{sec:Conclusion}
We have proposed a new UEP approach by partial superposition transmission using LDPC codes. The potential coding gain for the MID can be predicted by the discretized density evolution, which also shows that the performace loss is negligible for the LID. Simulation results verified our analysis and showed that, over both AWGN channels and uncorrelated Rayleigh fading channels, UEP-by-PST can provide higher coding gain for the MID compared with the traditional EEP approach, but with negligible performance loss for the LID. This is different from the traditional UEP approaches that usually degrade the performance of the LID while improving the performance of the MID. Simulation results also showed that UEP-by-PST is more efficient than UEP-by-Mapping in the DVB system from a practical perspective by taking as a criterion the minimum SNR required to satisfy simultaneously the error performance requirements for both the MID and the LID.

\section*{Acknowledgment}

The authors would like to thank Mr. Shancheng Zhao from Sun Yat-sen University for useful discussions.

\ifCLASSOPTIONcaptionsoff
  \newpage
\fi

\bibliographystyle{IEEEtran}


\balance
\end{document}